\documentclass[10pt,twocolumn]{article}

\newtheorem{theorem}{Theorem}[section]

\newtheorem{nt}{Selfnote}

\newtheorem{definition}[theorem]{Definition}

\newtheorem{remk}[theorem]{Remark}

\newtheorem{exmp}[theorem]{Example}

\def\FullBox{\hbox{\vrule width 8pt height 8pt depth 0pt}}

\def\qed{\ifmmode\qquad\FullBox\else{\unskip\nobreak\hfil
\penalty50\hskip1em\null\nobreak\hfil\FullBox
\parfillskip=0pt\finalhyphendemerits=0\endgraf}\fi}

\def\qedsketch{\ifmmode\Box\else{\unskip\nobreak\hfil
\penalty50\hskip1em\null\nobreak\hfil$\Box$
\parfillskip=0pt\finalhyphendemerits=0\endgraf}\fi}

\newfont{\inhead}{eufm10 scaled\magstep1}

\usepackage{amsfonts, amsmath,amssymb, graphics, setspace}
\usepackage{epsfig}
\RequirePackage{epsfig}

\usepackage{subfigure}

\newcommand{\fix}[1]{}
\newcommand{\look}[1]{}
\newcommand{\cut}[1]{}
\newcommand{\gnote}[1]{}

\begin{document}

\title{Potential Networks, Contagious Communities, and Understanding Social Network Structure}
\date{}
\author{
Grant Schoenebeck\thanks{University of Michigan, Division of Computer Science and Engineering, Ann Arbor, MI  48105,  e-mail: \texttt{schoeneb@umich.edu}. The author thanks the Simons foundation for their generous support of this research.}
}


\maketitle
\begin{abstract}
In this paper we study how the network of agents adopting a particular technology relates to the structure of the underlying network over which the technology adoption spreads.  We develop a model and show that the network of agents adopting a particular technology may have characteristics that differ significantly from the social network of agents over which the technology spreads.  For example, the network induced by a cascade may have a heavy-tailed degree distribution even if the original network does not.

This provides evidence that online social networks created by technology adoption over an underlying social network may look fundamentally different from social networks and indicates that using data from many online social networks may mislead us if we try to use it to directly infer the structure of social networks.  Our results provide an alternate explanation for certain properties repeatedly observed in data sets, for example: heavy-tailed degree distribution, network densification, shrinking diameter, and network community profile.  These properties could be caused by a sort of \emph{sampling bias} rather than by attributes of the underlying social structure.  By generating networks using cascades over traditional network models that do not themselves contain these properties, we can nevertheless reliably produce networks that contain all these properties.

An opportunity for interesting future research is developing new methods that correctly infer underlying network structure from data about a network that is generated via a cascade spread over the underlying network.

%
%

\end{abstract}


%

\section{Introduction}

The advent of Web 2.0 has tremendously enriched researchers' access to data.  Instead of observing eighteen monks for months waiting for something interesting to happen~\cite{Sampson69}, researchers now have access to approximately 160 million users' 90 million daily tweets through Twitter's API~\cite{Twitter}.   While these data tell us what people do online, it is less clear how much these data tell us about people in a broader context.

Social science researchers developed social networks as a methodological tool for understanding social phenomena, such as how  individuals' actions affect macro-level features of society, or how an individual's ``location" in a network affects his/her opportunities \cite{MarinW-09,Granovetter73}.  Sociologists have long distinguished between different types of networks \cite{MarinW-09}.  Some examples are trust networks: from whom would you feel comfortable asking for \$1000?; friendship networks: with whom do you want to go out Friday evening; information networks: with whom do you discuss important matters; and self-declared/articulated networks: who do you want the world to believe are your friends?

Social networks are not to be conflated with \emph{online social networks} such as LiveJournal, Epinions, MySpace, Facebook, and Twitter.  We will use the terms \emph{contagious networks} to denote networks that grow by adding new members where the new members are often ``infected" by their current social ties.  The key property of contagious networks is that people often join these networks because they have a friend or acquaintance that is already a member.  Contagious networks include most \emph{online social networks} because people are more likely to sign-up for such networks if they already have friends on them.  Citation networks, communications networks, collaboration networks, co-authorship networks, product co-purchasing networks may also be considered contagious networks.  Contagious networks provide much of the digital data we have about networks.  The actions of joining and participating in these networks (e.g. logging into LiveJournal, or coauthoring a paper) are often captured digitally.
Hence, contagious networks provide a means for studying social questions pertaining to social networks by providing data.

\vspace{5pt}

Because contagious networks are spread over an underlying social network, it is natural to conjecture that these networks share many properties.  However, it is difficult to know if this data generalizes past the digital world.   The importance of this distinction is indicated by a familiar question, ``who in the room is friends with his/her mother on Facebook?"  However, even if no one were Facebook friends with his/her mother would this meaningfully affect any large-scale measurements of the data?   Does the sheer scale of such data render differences between the contagious networks and social networks to be mere annoyances or do these differences present a substantial obstacle to using data from contagious networks to make inferences about social networks.  This is a key question that this paper hopes to address.

\subsection{Summary of Results and Implications}
We argue that the data from contagious networks is not tantamount to holding up a big mirror to our society; it is more like looking at our society in a fun-house mirror--where things may appear very differently than they are.

Using computer simulations, we illustrate examples where the contagious network and the underlying network have very different properties.
Data mining has shown that many contagious networks share a few common features:  heavy-tailed degree distributions, shrinking diameters, edge densification, and a particular ``network community profile".  We show, with computer simulations, that even though certain well-known network models (e.g. the Watts-Strogatz model or a collection of cliques) have none of these properties, if we use these models as an underlying network and grow contagious networks over them in a natural way, then the resulting contagious networks have all of these properties.  We investigate various models of transmission and show that these results are robust to changes in the model.  We study various parameter regimes to understand when our results hold.  We also explore the theoretical mechanisms underlying our experimental results.  In the case of degree distribution, we can prove that certain underlying structure will endow the contagious networks with heavy-tailed degree distributions even when the underlying network is regular.

While these models are admittedly stylized, we believe that they are natural, and that these results give strong evidence of important implications, which we summarize here; they are discussed in more detail in Section~\ref{'implications'}.

%

\vspace{8pt}


(1)~These results provide a natural framework for developing generative models for contagious communities which capture the aforementioned four properties: start with a model for an underlying social network and model a contagion spreading over it.
(2)~These results provide strong intuition that we need different models for social networks and contagious networks.  It may be a mistake to import social network intuition into models for contagious networks.  Similarly, by datamining contagious networks, one expects to find attributes that are common amongst contagious networks; however, these observations may not apply to the underlying social network.
(3)~This helps make sense of the counter-intuitive results of Leskovec et al~\cite{LeskovecLDM08-jounal} about network structure.  These results make intuitive sense in context of contagious communities, but may not apply to other social networks.   More speculatively, if we imagine these networks as being a community, it may allow Leskovec et al to give us insight into the structure of communities as well.  (4) In a model where social networks are not created \emph{ex nihilo}, but from existing social structures, contagious  networks provide a sort of sampling technique for learning the underlying social network.  While it may be impossible to directly infer the underlying social structure, more subtle techniques might work.
In the Section~\ref{'section-potential networks'}, we pose the question: if contagious network data is akin to looking in a fun-house mirror, then what aspects of reality can we still reliably deduce from looking at this data?  After all, if you see feet and a head in a fun-house mirror, you can be fairly certain that there is a body in between.

\subsection{Related Work}
Technology adoption as a process on a social network has been studied and documented before; however, usually only the size of the cascade is considered.
For example, in an experimental study, Centola~\cite{Centola10} creates online communities populated with volunteers and studies the spread of joining a health forum network over this strictly enforced underlying network.  Centola was mostly concerned with what types of underlying network structures would foster the largest cascade.  For more examples, see Chapters 6 and 9 of~\cite{Jackson08}.

Here, we model on-line social network formation as technology adoption, and investigate how the network \emph{structure} of the on-line social network is affected by the underlying social network structure. In fact, we condition on the size and so explicitly remove this variable from study.

This phenomenon of network creation over existing structure extends to many settings beyond on-line communities.  Segal~\cite{Segal74} observes that the best prediction of who would become friends at a certain police academy was the proximity of their last names in the alphabet (this was presumably due to the frequent placement of the cadets in alphabetical order).  Thus the last names indicated a certain underlying social structure over which friendships eventually formed.  In another study, also at a police academy, Conti and Doreian~\cite{ContiD10} show that seating assignments and squad assignments  predict friendship ties.
In this case, the study tries to manipulate the underlying social structure networks to foster inter-racial camaraderie at the academy.

\vspace{8pt}

Our results can be understood as trying to study the sampling bias resulting from the technology adoption process.  The same person may interact over many differ types of technology--telephone, text, email, Facebook, Twitter--or without technology.  Each technology may be \emph{selectively} used for particular communication needs.  Each user may use several distinct instances of the same technology (e.g. a work and home telephone, several email accounts).  To use this data to make assertions about social questions, we need to know that the data generalizes past the digital world, at least in the cases that we care about.

A series of work (e.g.~\cite{LakhinaBCX03,AchlioptasCKKM-05}) points out a similar sampling bias in context of traceroute sampling.  For example,  Achlioptas, Clauset, Kempe, and Moore~\cite{AchlioptasCKKM-05} show that traceroute sampling finds power-law degree distributions even in regular random graphs (which are very far from having a power-law degree distribution).  A sampling bias caused by using traceroute sampling means a power-law distribution can be measured even when the underlying degree distribution is constant.

However, traceroute sampling is fundamentally different than the cascading processes evaluated here.\footnote{Trace-route looks at the degree distribution on a breadth first search tree.  In our setting, this would be similar to $RET(n, \alpha = 0, \beta = 1)$--every infected node immediate infecting all uninfected neighbors but no internal edges which, as we will see, is outside the parameters that yield our results or that are interesting in our setting.}  In particular, unlike in the traceroute sampling case, running our models of cascades over Erd\"os-R\'enyi random graphs does not yield power-law or heavy-tailed degree distributions.  Thus it must be a different mechanism acting in each case.

\paragraph{Terminology}

For expositional convenience and concreteness, throughout this paper we will use off-line friendship interchangeably with social network and as the canonical example of a social network.  Likewise, we will use on-line social networks and sometimes Live Journal\footnote{LiveJournal is an early blogging and social networking community.} in particular to be a stand-in for contagious networks.  A sharp distinction between contagious and social networks is not always clear, but nonetheless we believe these generalizations are useful.  Also, not all on-line social networks are necessarily contagious communities as membership in some may not be spread primarily via an underlying social structure.  While the language of cascades and adoptions is more accurate and traditional to describe the spread of some cultural artifact, we will interchangeably use the notation of a virus and infection because such terminology is it often more concise.

\paragraph{Road Map} Section~\ref{'Section Models'} describes the models we use to construct the underlying social structure,  the processes by which contagious networks spread over this structure, and the properties we are interested in comparing between the original network and the contagious network.  The results of simulations over these models are summarized in Section~\ref{'section Simulation Results'}.  Section~\ref{'section Theoretical Models'} presents some theoretical rational for these results.  In Section~\ref{'implications'} we draw implications of our results.  Finally, Section~\ref{'section-potential networks'} concludes with what we feel is an interesting open question raised by our study as well as a framework with which to approach it.

\section{Models and Formalisms} \label{'Section Models'}

In this section we present natural models of underlying social networks and cascades which spread over these networks to create contagious networks.  We give examples of properties found across many different network data sets.  In subsequent sections we will start with one of these network models, simulate the growth of a contagious network over it, and then compare properties of the contagious network to those of the underlying social network.

\subsection{Graph Models}
 \label{'subsection graph models'}
For our underlying social networks, we use simple and traditional generative models which do not exhibit the characteristics we are hoping to capture.
The two graphs that we focus on for our potential networks are the Watts-Strogatz model and the Planted Community model.  Each is characterized by two properties: 1) ``random" short cut paths, and 2) edges that provide a lot of clustering but generally fail to provide shortcut paths.

The \emph{Watts-Strogatz random network model} is defined by three parameters.  The undirected $WS(n, d, r)$ ensemble of random graphs--where $n$ is the number of vertices, $d$ is the average degree which is even, and $r \in [0, 1]$ is a parameter--are defined by the random process that creates them.  This process begins with the graph on $n$ nodes $\{0, 1, \ldots, n-1\}$ where each node is connected to the $d$ closest other nodes so that $E = \{(k, k\pm \ell \mod n): 1 \leq \ell \leq d/2\}$.  With probability $r$, each edge $(u, v)$ is then ``rewired", that is replaced with the edge $(u, v')$ where $v'$ is chosen from the vertices not already connected to $u$.\footnote{The original WS definition is slightly more complicated than this because the order which you consider the edges may matter, see \cite{WattsS98} for the details of ordering.  We use the implementation in SNAP~\cite{SNAP10} which ignores these subtleties.}

The \emph{Planted Community Model} $PC(n, d, r)$ is defined by the same three parameters as the Watts-Strogatz model but here we require that $n$ is a multiple of $d$.  To create such a graph, the vertices are partitioned into $n/d$ equally sized cliques of size $d$.  Each edge $(u, v)$ is then ``rewired" with probability $r$.

\paragraph{Models of Transmission} \label{'subsection cascade models'}
In this section we first define four simple models of transmission.

The first, which we call \emph{random edge transmission induced graph}, has one parameter.  $RETIG_G(m)$ is defined by starting with the graph $G = (V_G, E_G)$ and initializing the infected set $I \subseteq V_G$ to a single random vertex.  A random edge $(u, v)$ is chosen uniformly from $E(I, \bar{I})$ and the vertex $v$ is added to $I$.  This is repeated until $|I| = m$.  The resulting infected graph is $G(I)$, the induced subgraph of $G$ on the vertices in $I$.

The first model includes all the edges in $G$ between vertices that are in $I$.  In the second model, these edges must also be discovered.  We call the second model  \emph{random edge transmission}, and it has three parameters.  $RET_G(m, \alpha, \beta)$ is defined by initializing the infected graph $H = (V_H, E_H)$ to the graph $(\{v_0\}, \emptyset)$ where $v_0$ is a random vertex from the potential graph $G = (V_G, E_G)$.   At each step, each edge $(u, v) \in E_G(V_H, V_H) - E_H$ is added to $H$ with probability $\alpha$ and each edge $(u, v) \in E_G(V_H, \overline{V_H})$ is added to $H$ (along with $v$) with probability $\beta$.  The process is run until $m$ additional vertices are included.

The third model  \emph{random edge transmission with multiple initial vertices} $RETMIV_G(m, \alpha, \beta, s)$ is defined the same way as $RET_G(m, \alpha, \beta)$ but the transmission is started from $s$ random vertices simultaneously.

Note that $RETIG_G(m)$,  $RET_G(m, \alpha, \beta)$, and \\$RETMIV_G(m, \alpha, \beta, s)$ never add edges that are not in $G$.

We create a more complex model which allows people in ``infected" communities to make new friends within the cascade.
The \emph{random edge transmission with exploration}  $RETWE_G(m, \alpha, \beta, \gamma)$ is defined exactly like the \emph{random edge transmission} except that at each round, for each triple $u, w, v \in V_H$ where $(u, w), (w, v) \in E_H$ the edge $(u, v)$ is added to $E_H$  with probability $\gamma$ (this edge is added with probability $\gamma$ for each such triple).



\subsection{Properties studied}

\paragraph{Heavy-tailed degree distributions}  Previous research has shown that many networks have heavy-tailed degree distributions\cite{AmaralSBS2000,BarabasiA99}.  By \emph{heavy-tailed degree distributions} we mean that the degree distribution approximates a straight line when plotted with both axes logarithmically scaled, perhaps followed by a drop-off.\footnote{While this terminology is not standard, we use because it captures the operational definition in many other papers,  provides enough precision to describe our results,  and avoids the controversy of the term power-law (see discussion on page 60 of \cite{Jackson08}).}  Power-law distributions, the related Yule distributions, and truncated power-law distributions are all heavy-tailed distributions~\cite{DraiefM-09}.  Often heavy-tailed degree distributions serve as a contrast to Poisson distributions, which are much more highly concentrated and have a much thinner tail (fewer points far from the average~\cite{Jackson08}).


\paragraph{Shrinking Diameters and Edge Densification}
Previous research has also shown that, over time, the diameter of contagious networks tends to shrink and that the average degree of vertices tends to increase
~\cite{LeskovecKF05}.
This work was based on analyzing four networks: the ArXiv citation graphs (for high-energy physics theory), the U.S. Patent citation graph, the graph of routers of the Internet, and the ArXiv affiliation graph (on certain topics).  Note, however, that none of these is actually an online social network in the usual sense.

\paragraph{Network Community Profile}

Another network feature that we are interested in is called the \emph{network community profile} and is described by Leskovec, Lang, Dasgupta, and Mahoney~\cite{LeskovecLDM08-jounal}.   The authors develop a tool to analyze network structure that they call the ``network community profile"-which we will describe shortly.  They show that this tool yields similar results when applied to over 70 data sets, such as LiveJournal.  In particular, the network community profile on the on-line social networks: LiveJournal, Epinions, LinkedIn, Del.icio.us, and Flickr look nearly identical (see \cite{LeskovecLDM08-jounal} pages 22 and 25).
They note that the plot decreases until around 100, then it stays roughly even for a short period, and finally starts to increase.
Finally, they show that this tool yields completely different results on virtually all generative models (except for one that they call the Forest-Fire model).

Leskovec et al were interested in studying the community structure on networks.  They define a community as a set of nodes with low conductance--many edges within the set compared to the number of edges leaving the set.  Even in very large datasets of contagious networks, they found few large communities (over 100 people) that fit this definition.  Broadly speaking, they found that the structure of these graphs was composed of ``whiskers" and a ``core".  \emph{Whiskers} are a set of nodes connected to the rest of the graph by only a one or a few edges.  The \emph{core} is a big connected tangle with no subsets of small conductance.  The ``community" structure that they detect (sets with low conductance) can be almost entirely attributed to collections of whiskers--groups just barely connected to the rest of the graph.

The \emph{conductance} of a set denoted $S \subseteq V$  $$\Phi(S) = \frac{E(S, \bar{S})}{min\{degree(S), degree(\bar{S})\}}$$ is equal to the number of edges leaving a set $S$ divided by the sum of the degree of the vertices in $S$ (or $\bar{S}$, whichever is smaller).  Thus, if $S$ is insular and does not have many edges leaving it relative to its total degree, then $S$ has low conductance.  The community network profile finds the set of each size $s: 1 \leq s \leq |V|/2$ with the lowest conductance and then plots this graph.  I.e $f_G(x) = \min_{S: |S|=x} \Phi(S)$.

\gnote{You may think about cutting the following paragraph}
The network community profile is closely related to isoperimetric inequalities which are a mathematical subject concerned with minimizing the boundary for a given volume (such as the circle in the plane), and thus showing that a large volume implies a certain sized boundary.  Here the ``volume" corresponds to the total degree and the ``boundary" corresponds to the number of edges leaving.

\vspace{10pt}

\section{Simulation Results} \label{'section Simulation Results'}

In this section we will describe the results of our simulations.  We compare the studied properties on the contagious networks and the underlying networks.
Overall, the simulations support our theory that contagious networks look like a cascade across simple network formation models.

Our results mainly apply to the beginning of a cascade.  Once the cascade reaches the entire graph, then by definition the underlying graph and cascade look the same.

 All simulations were done using the SNAP System~\cite{SNAP10}.  This is particularly important for the network community profile and diameter which are both approximated with heuristics.  These heuristics were shown to work well in other graphs (see Section 5 of Leskovec et al~\cite{LeskovecLDM08-jounal}), but that is no guarantee that they work well here.  \footnote{The computer code used is available on the author's homepage.}

Unless specified otherwise, simulations were run on an underlying graph of 1,000,000 nodes with average degree 100, using a rewiring parameter of $0.1$ and the random edge transmission models with $\alpha = 0.7$, $\beta = 0.01$ and $\gamma = 0.001$.

We first describe the results when the Watts-Strogatz model is used as a potential graph.  We study four properties of the contagious network: degree distribution, diameter, edge densification, and network community profile.

\paragraph{Degree Distribution}

%
%

\begin{figure*}
  \centering
  \mbox{
    \subfigure[Watts Strogatz graph\label{degrees2}]{\includegraphics[width = 3.4in]{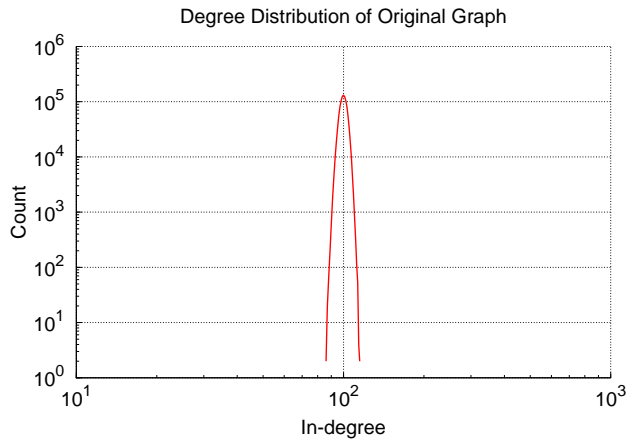}}\quad
    \subfigure[Cascades on Watts Strogatz graph after 625; 5,000; and 80,000 nodes infected.  Guidelines indicate a power-law with exponent -1.11\label{degrees1}]{\includegraphics[width = 3.4in]{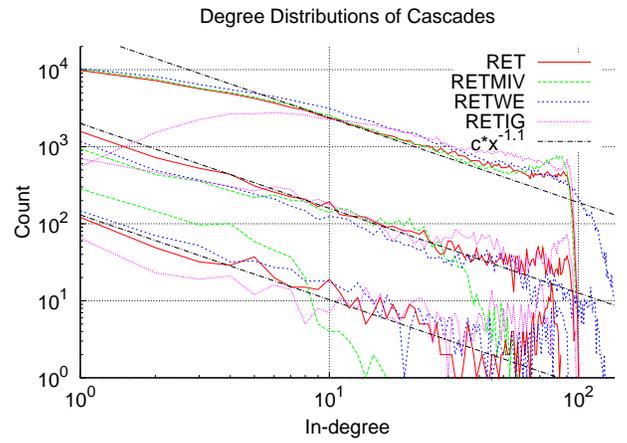}}
  }
  \caption{Degree Distribution}
  \label{degrees3}
\end{figure*}

%

Even though the degree distribution of the Watts-Strogratz model is highly concentrated, the resulting contagious networks have a degree distribution that resembles a heavy-tailed distribution (see Figure~\ref{degrees3}).  This is especially surprising since the maximum degree of the original graph $G$ (and hence the largest possible degree in $H$) was slightly over 100 in the trials we ran. Eventually, the degree distribution looks like a truncated heavy-tailed distribution.  That is, all the points present are in a straight line, but this tail suddenly stops when the underlying graph has no vertices with degree above a certain value.

After a large fraction of the underlying network becomes infected, we expect this effect to go away because the underlying networks does not have a heavy-tailed degree distribution.  These plots hold to approximately straight lines until the cascade reaches 80,000 nodes at which point they started to diverge.  After about 1/3 of the graph is infected, these plots begin to diverge substantially from the truncated heavy-tailed distribution.  The same behavior was observed for all transmission types.  However, if $\alpha$ is made too low, or $\beta$ too high, then this behavior is not as prevalent. The extreme case where $\alpha = 0$ and $\beta = 1$, is very similar to the trace-route sampling setting.  On regular graphs the contagious networks limit to a power-law degree distribution~\cite{AchlioptasCKKM-05}, but the underlying graphs that we consider do not yield a contagious network with a heavy-tailed degree distribution for these settings of the parameters.


%
%
%

\paragraph{Diameter and Edge density}


\begin{figure}
\centering
\includegraphics[width=3.4in]{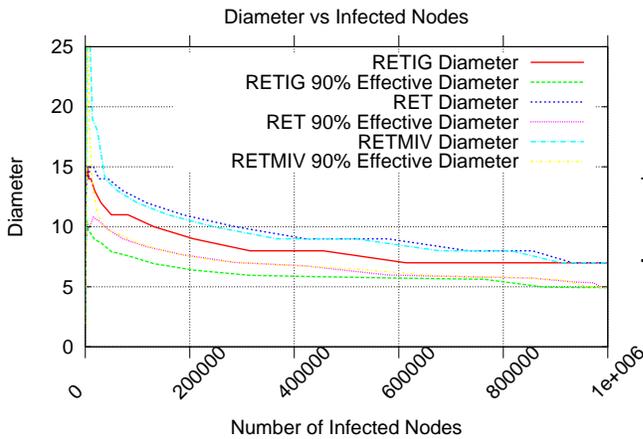}
\caption{Diameter and 90\% effective diameter as a cascade spreads on Watts-Strogatz graph.} \label{'diameter'}
\end{figure}

\begin{figure}
\centering
\includegraphics[width=3.4in]{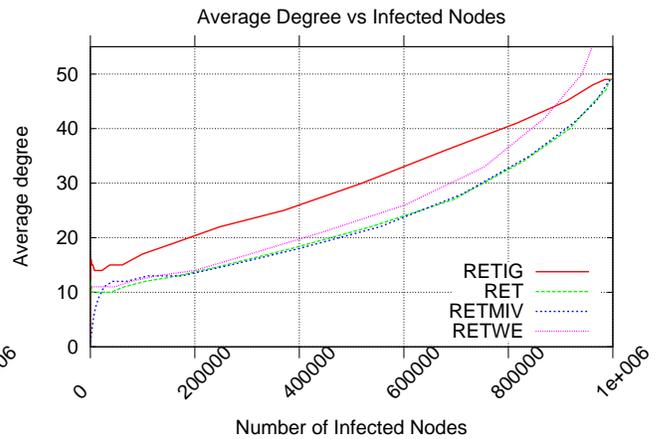}
\caption{Average degree vs infected Nodes as a cascade spreads on Watts-Strogatz graph.} \label{'avgdegree'}
\end{figure}

We also observe the diameter and average degree of the network.  We find the diameter and effective diameter shrink and the average degree increases in accordance with the results of Leskovec, Kleinberg, and Faloutsos~\cite{LeskovecKF05} (see Figures~\ref{'diameter'}~and~\ref{'avgdegree'}).  Both plots qualitatively match the plots in \cite{LeskovecKF05}.  The edge density increases approximately linearly, after an initial jump.  The diameter, after a large spike and ensuing drop off, decreases gradually.

\begin{figure*}
  \centering
  \mbox{
    \subfigure[Watts-Strogatz Graph \label{WS-ncp-plot0}]{\includegraphics[width = 3.4in]{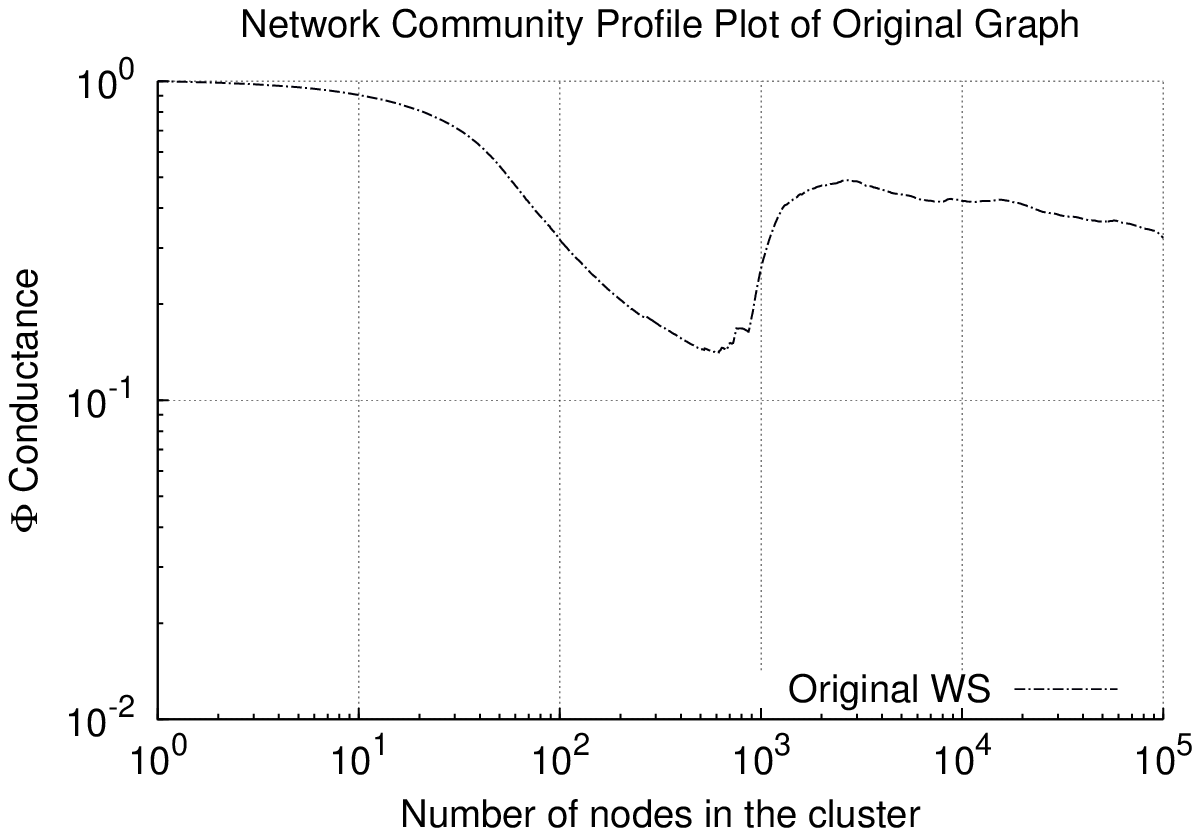}}\quad
    \subfigure[Cascades on Watts-Strogatz Graph after 80,000 nodes infected . \label{WS-ncp-plot}]{\includegraphics[width = 3.4in]{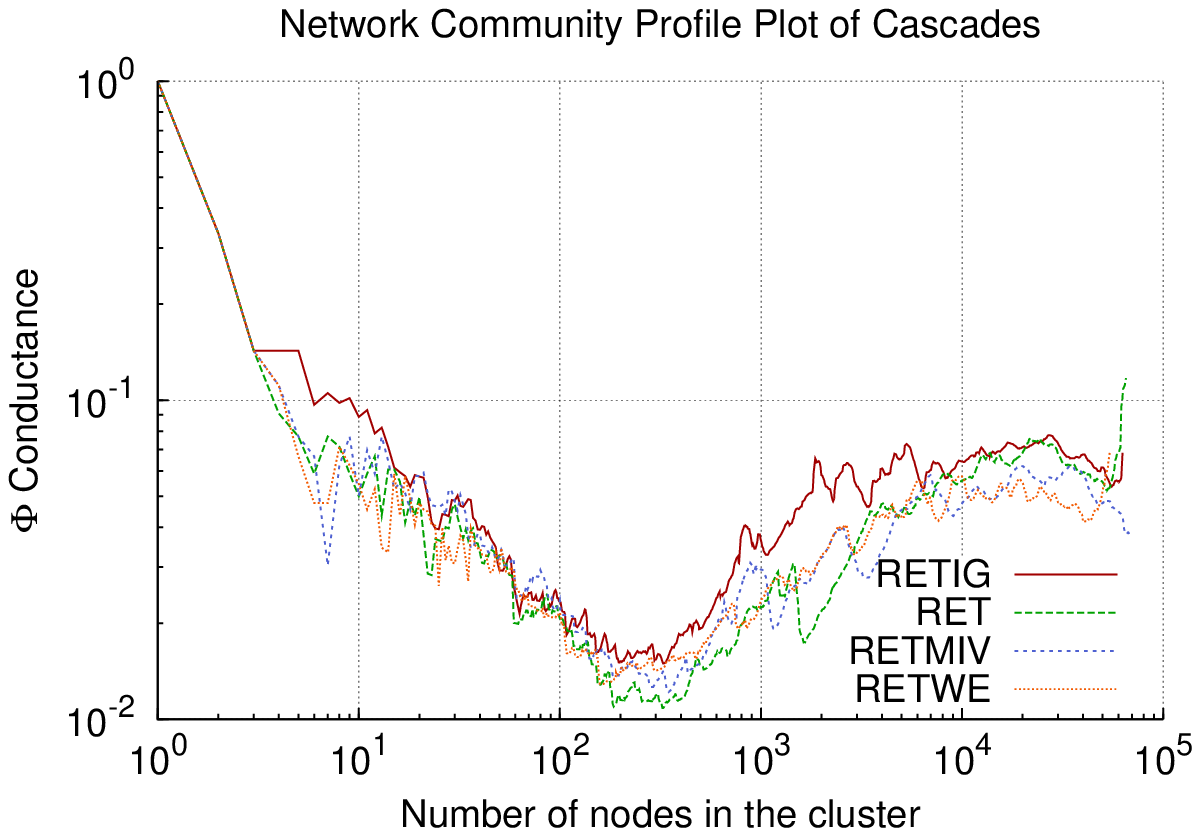}}
  }
  \caption{Network Community Profile}
  \label{NCPfigure}
\end{figure*}

%

\paragraph{Network Community Profile}
We found that the network community profile closely matches that of the online social networks that Leskovec et al studied in \cite{LeskovecLDM08-jounal} in all our models of transmission.   Figure~\ref{NCPfigure} shows both the original network community profile of the Watts-Strogatz model and the network community profile of a virus spread over the network.  This similarity holds up until about 1/3 of the vertices in the graph are infected.  Differing the population size and degree does not seem to affect the outcome.  However, if the rewiring probability $r$ is made too large ($> .3$) then the shape collapses; the plot never decreases sufficiently.   A similar pattern occurs if $\alpha$ is not sufficiently large compared to $\beta$; the edges between nodes of the infected graph $H$ fail to fill in and no community structure is detected.



\paragraph{Other Graph Generation Models}
The results for the Planted Community model are nearly identical to those of the Watts-Strogatz model, which implies a certain robustness of these results.

Our hypothesis was not confirmed on all graphs.  We do not observe all of these behaviors when we run these processes on various graph generation models including Erd\"os-R\'enyi random graphs~\cite{ErdosR60}, Preferential Attachment networks~\cite{BarabasiA99}, or complete graphs.  We hypothesize that, in the Erd\"os-R\'enyi random graphs and the Preferential Attachment model, this is because there is very little clustering to begin with, causing the virus to spread evenly over the graph in a tree like fashion and remain unclustered.  Such behavior might not continue if nodes ``met" other infected nodes by virtue of being infected and having a common neighbor.  To test this hypothesis we embellish the dynamics to artificially add community structure using $RETWE$ as a model of spreading.  We find that the network community profile still does not look like the sought after behavior, though it comes closer.

In the complete graph, we hypothesize that the reason contagious communities do not contain all these properties is that there are no ``short-cut" edges.


\section{Theoretical Insights} \label{'section Theoretical Models'}

In this section, we present mathematical insights that elucidate many of the empirical results of the previous section.  In particular, we show that a graph exhibiting both ``strong" and ``weak" ties should generate contagious networks exhibiting a heavy-tailed degree distribution.  This theory accurately predicts the results in the aforementioned section.

The RETIG model is mathematically identical to the model of \emph{first passage percolation}\footnote{In first passage percolation typically one node starts infected.  Each edge is equipped with a clock, and the infection can only travel across each edge when it rings} where each edge is equipped with a Poisson clock of unit rate,  the number of vertices is conditioned on, and all induced edges are included.  First Passage percolation has been studied on Erd\"os-R\'enyi random graphs, but to our knowledge, not on graphs with small world properties~\cite{VanDerHofstadHV-01}.

While all these properties: degree distribution, diameter, network community profile (referred to as isoperimetric inequalities in this literature), and density have been studied on certain graphs and even in certain percolation models, we know of no results that apply directly to the situation at hand.  In particular, site and bond percolation has been studied (in site percolation each node is present/removed with some independent probability and in bond percolation each edge is present/removed with some independent probability).  However, these models differ substantially from the contagion model.  Moreover, these results tend to focus on the property of component size, which we fix a priori.

\paragraph{Degree Distribution}

We start with a theorem about a family of graphs closely related to the planted community model.

\begin{definition}  The \emph{planted clique model}  $PCM(n, k, r)$ generates a graph on $n$ nodes by superimposing the edges from a  degree $rk$ regular random graph and from partitioning the nodes into $n/k$ cliques of $k$ nodes each.  \end{definition}

Next we show that RETIG will produce a power-law degree distribution on such graphs.

\begin{definition}  A power-law distribution with exponent $-\gamma$ is a distribution on the positive integers where $p(x) \propto L(x)x^{-\gamma}$ such that $\lim_{x \rightarrow \infty} L(tx)/L(x) = 1$.   \end{definition}

\begin{theorem}  Let $G$ be a family of $PCM(n, \sqrt{n}, r)$ graphs where $0< r < 1$ is a constant.  Then as $n$ increases, the degree distribution of the $RETIG$ on $G$ after infecting $\sqrt[4]{n}$ nodes will limit to a distribution $D$ which is a power-law distribution with exponent $-1 - r$.  \end{theorem}


The proof follows from the following more general intuition about the Yule distribution~\cite{Yule-25,DraiefM-09,Newman-05}.  The Yule distribution was created to model the following situation concerning species and genera:  at each time step choose a random species and with probability $1-\alpha$ the species creates a new species in the same genus, and with probability $\alpha$ the species creates a new species in a new genus. The Yule distribution describes the fraction of genera with a particular number of species in the limit of this process.  The tail of this distribution limits to a power-law distribution with exponent $-2-\frac{\alpha}{1-\alpha}$.

In the $PCM$ graphs, in the limit of $n$, a  $\frac{1}{1 + r}$ fraction of each vertices neighbors are \emph{cliquish}--in the same clique--and a $\frac{r}{1 + r}$ fraction of its neighbors are \emph{distant}--in different cliques, i.e. neighbors via the random edges.  Let $\gamma = \frac{r}{1 + r}$.  Think of $r$ being small so that $\gamma$ is close to 0, and consider a cascade over such a network. If we assume that the cliques are sufficiently large, that there are sufficiently many, and that the cascade has not been going too long (so that each vertex has about the same number of non-infected neighbors), then when a new vertex is infected, it is like picking a random vertex in the cascade and infecting a random neighbor.   This neighbor has a $1-\gamma$ probability of being cliquish (in the same clique) and a $\gamma$ probability of being distant (in a different clique).   Because of this the number of nodes present in each clique closely follows a Yule distribution with parameter $\gamma$, and thus limits to a power-law distribution with exponent $-2 - \frac{\gamma}{1 - \gamma} = -2-r$.

Consider the induced subgraph of the cascade.  The degree of each vertex will be equal to the number of cliquish and distant neighbors that are also included.  The number of cliquish neighbors of a vertex is simply equal to the number of infected nodes in its clique.

Because the number of infected vertices in each clique limits toward a power-law distribution with exponent $-2-r$ and a clique with $k$ infected vertices contains $k$ vertices of cliquish degree $k-1$, the degree distribution of cliquish edges limits toward a power-law distribution with exponent about $-2-r + 1$.  Thus the degree distribution will be a power-law with exponent $-1-r$.  It turns out that distant neighbors contribute very little to a vertex's degree in comparison  to the cliquish neighbors because the distant neighbors form a random graph.  The theorem follows from the above intuition.

Thus in Section~\ref{'section Simulation Results'} we expect that in the RETIG model we have a power-law with exponent $\approx -1.1$.  We see in Figure~\ref{degrees3} of  Section~\ref{'section Simulation Results'} that the data fits this well.

Several things break down after the cascade continues to spread.  First some vertices in the cascade may have a significant fraction of their neighbors in the cascade.  Alternatively, some clusters may have a significant fraction of vertices in the cascade.  This means that these vertices (or clusters) are less likely than random to spread to a friend (or gain a new adoptive member).  Secondly, as the cascade infects a significant number of nodes in the graph, a random edge may not spread the virus to a new area of the graph, but instead may reach an already infected area.

While the Watts-Strogatz graph with rewire parameter $r$ is not a collection of cliques, it behaves in a similar manner to the above graph.  Locally, it looks a lot like a clique.  If one imagines several locations of the Watts-Strogatz graph being infected, then each location is ``clique-like" in that most vertices in that location are neighbors.  When a new vertex is included, with probability $r$ the link will be a rewired link and thus is likely to start a new location of the infection.  However, if an original (non-rewired) edge is included, then the new vertex will be in a similar location to previously included vertices (and thus is likely neighbors with most of them).

This argument uses two properties: that networks are locally clique like, and that a fraction of the edges are random.  In general, when an underlying network has both high clustering and an $\alpha$ fraction of its edges are ``random", we expect the degree distribution of the contagious network to look like a modified Yule distribution for the same reasons.


\paragraph{Edge Densification}

While it seems intuitive that the edges will densify in a cascade, it turns out that the schedule of densification differs between an Erd\"os-R\'enyi graph and a graph with clustering.

The edge density of a cascade on an Erd\"os-R\'enyi graph will be proportional to the fraction of the network infected by the cascade.  Thus, the cascade does not densify until it reaches a constant fraction of the graph.  With high probability, all subsets of a small constant fraction, say $\gamma n$, of nodes in an Erd\"os-R\'enyi graph will have average degree less than $2+ \epsilon_{\gamma}$ where $\epsilon_{\gamma}$ depends on $\gamma$ and goes to 0 as $\gamma$ does.  (See, for example, the appendix of \cite{S08}).  Thus densification cannot happen in a cascade on an Erd\"os-R\'enyi random graph with expected constant degree until it infects a significant fraction of nodes.  However, as we previously saw, we expect the degree distribution of the cascade in a Watt-Strogatz graph with rewiring probability $r$  to emulate a power-law distribution with exponent $-1 -r$ even when only $\sqrt{n}$ vertices have been infected.  Thus for $r< 1$ the expected edge density is infinite. Recall that the Yule distribution is what is expected in the limit, so we expect the edge density to grow even before reaching a constant fraction of the vertices.

\paragraph{Forest Fire Model as a Contagious Community}
\label{'ForestFireEquivalence'}
Leskovec et al found that the Forest Fire model was the one generative model they tested that did replicate the results of the community network profile that they found on the 72 data sets.  We note that the exploration component of $RETWE$--that is, adding direct links to neighbors of neighbors--run on a random graph intuitively simulates the Forest Fire model.

The complete Forest Fire model can be found on page 9 of Leskovec, Kleinberg, and Faloutsos~\cite{LeskovecKF05}.  For our purposes, it will suffice to present a slightly simplified undirected version. Our Forest Fire model has one parameter $p$, the burning probability.  The model starts with a single node.  At each time step a new node $v$ joins, chooses an existing node $u$ at random, and forms a link with $u$.  For each node $w$ that $v$ links to (starting with $u$), $v$ also links to $k_w$ of $w$'s neighbors where $k_w$ is chosen from a binomial distribution with mean $(1-p)^{-1}$. This is guaranteed to terminate because $v$ is not allowed to link to any node more than once.

Consider running $RETWE$ on a low degree Erd\"os-R\'enyi random graph.   When a vertex $v$ joins (if the contagious network has not reached more than a small fraction of the total nodes), then it is very likely that $v$ is attached to exactly one node $u$ of the infected subgraph, $H$, (the vertex that caused $v$'s infection).  The vertex $v$ can add more ties in the infected subgraph $H$ by ``exploration" on the infected subgraph through ties of $u$ in $H$ (that is adding direct ties to neighbors of $u$).  Each time that $v$ links to a neighbor $w$ of $u$ in $H$, the next time step $v$ can add nodes to neighbors of $u$ as well.

The difference between these two models is in the number of neighbors that $u$ finds by exploration.  In the forest fire model it is $(1-p)^{-1}$ in expectation, and in the $RETWE$ model it depends on the amount of time the node has been in the network.   Also in $RETWE$ a vertex can additionally add ties by infecting neighbors in $G$ that are not yet in $H$.

Thus it is not surprising that both of these models produce similar though not certainly not identical network community profiles.

\section{Implications} \label{'implications'}

We think that there are several important implications from the above models and simulations.

\noindent\textbf{New generative model for contagious networks:}
This intuition provides us with a new generative model for contagious networks.  Start with a social network model, and model a contagion spreading over it.
We show that with certain modeling choices (for example Watts-Strogatz with RET adaption) this two-step simulation captures both the intuition of sociology research about social network models--small diameter~\cite{Milgram67} and local clustering~\cite{Watts99}--and the datamining research on contagious networks--shrinking diameter and edge densification~\cite{LeskovecKF05}, heavy-tailed degree distribution~\cite{BarabasiA99}, and a particular network community profile~\cite{LeskovecBKT08}--all in one simple and intuitive model.  We acknowledge that our starting networks (e.g. Watts-Strogatz) are very stylized and not particularly realistic, and we leave it for future work to further develop this framework with more realistic underlying networks and adoption patterns.

\noindent \textbf{Contagious networks and social networks require different models:}
We show that metrics that appear to test global properties (e.g network community profile) and metrics that appear to test local properties (e.g. degree distribution) may show dramatically different results on contagious networks and the underlying social networks.  While this observation has been made before, we provide results that begin to show the scope and scale of the qualitative and quantitative differences.

A long line of work seeks to study network generation models (for examples see~\cite{ErdosR60,BenderC78,FrankS-86,WattsS98,BarabasiA99,LeskovecKF05,LeskovecBKT08,LattanziS-09}).
Our results warn that it is unlikely that any one model will serve to generate realistic models for broad class of social networks.  This remains true even if we only desire that our models capture fairly basic properties.
Indeed, we should not \emph{a priori} expect all properties to be universal, and thus we should not \emph{a priori} expect one generative model.  One of the original motivations for sociologists to develop social network theory was to explain how social networks \emph{differ} and to understand the implications of these differences.  For example Gans \cite{Gans62} studied how Boston's West End community was unable to form a coalition to fight an ``urban renewal" measure that ended up destroying the community, even though other seemingly similar communities were able to organize against and defeat such measures~\cite{Granovetter73} and, moreover, how social structures could have contributed to this outcome.

Distinguishing the two tasks of modeling contagious networks and social networks gives a partial explanation for the difficulty in the task of creating realistic models.  By not distinguishing the two tasks, on the one hand, social networks intuition is inadvertently imported into models of contagious networks.  However, this intuition is found to be incorrect by datamining contagious communities.  On the other hand, the counter-intuitive findings of datamining contagious networks is being advertently brought into social network models, where it makes little intuitive sense.

If indeed social networks and contagious networks are different, this indicates that using data from contagious social networks may mislead us if we try to directly use it to understand social networks.
There is selection bias toward datamining contagious networks because data is more easily available for this type of network.  Thus, we would expect datamining studies to find attributes that are common amongst contagious networks, but not necessarily present in social networks.  Yet, despite the prevalence of certain characteristics (such as heavy tail degree distribution \cite{BarabasiA99}), models without these characteristics may still be valid in a wide variety of interesting settings.  In particular, it may be that such characteristics are common to contagious networks, but are not found in many social networks.  While we cannot show that contagious networks are necessarily different from social networks on all these metrics, we do remark that if the intuition that guided the first generative models is correct (which do not contain heavy-tailed degree distribution, shrinking diameter~\cite{LeskovecKF05}, edge densification, and a particular network community profile~\cite{LeskovecBKT08}) then such a discrepancy must exist.  The counter-intuitive core and whisker structure found by~\cite{LeskovecLDM08-jounal} might accurately characterize actual social networks; however, we feel that contagious networks is a more natural explanation of these observations.

At the same time, this distinction frees us from conforming to the intuition of sociologists when modeling contagious communities.  Leskovec et al~\cite{LeskovecBKT08}, already showed us that contagious networks are not as we would commonly believe them to be.  Some doubted the counter-intuitive results.   Our experiments provide intuition that supports their observations and show that natural cascades will lead to heavy-tailed degree distributions, edge densification, shrinking diameter, and certain network community profiles  in contagious networks.

More speculatively, this gives us an opportunity to \emph{re-imagine what a community is and what they look like} through the results of Leskovec et al.  These contagious social networks can be seen as a \emph{community} within an underlying social network.  That is the nodes of LiveJournal form the ``LiveJournal community'' which is embedded in society.  The LiveJournal network can be viewed both as a network in and of itself, but also as a community in a larger network.  We can perhaps use the core/whisker model of Leskovec et al to understand properties of communities.

This model provides an alternate view of community structure compared to that metrics such as modularity~\cite{NewmanG04} and conductance~\cite{LeskovecLDM-08}.  This view sees communities as gradually adding internal connections and external members who start on the periphery of the group but gradually gradually become more central to the community.  Communities are composed of whiskers and a core that has no insular communities.


If dynamics similar to our model exist, then contagious social networks will, with high probability, contain properties for a range of underlying social networks even though these same underlying networks may or may not have these properties.  We note that our results only hold when the network is fraction of the total graph.  Thus we think they would apply more to  LiveJournal than Facebook.  Additionally, as we already remarked, not all on-line social networks are necessarily contagious.  For example, we do not expect the link structure of anonymous on-line support groups to be generated by a cascade over an underlying social structure.


It would be interesting to go beyond simulation data and attempt to verify this distinction between social networks and contagious networks on real data.  However, it is not entirely clear what data set would be a good test.  To a certain extent, this is really asking the impossible.  How does one accurately measure a ``trust" network (even between two people)?  However, even in a limited context, it would be interesting to carry out such a study.

\subsection{Opportunities}

At the same time, these results point toward the opportunities (and challenges) of developing techniques for reconstructing the underlying social network from contagious network data.  In a model where social networks are not created \emph{ex nihilo}, but from existing social structures, contagious  networks provide a sort of sampling technique for learning the underlying social network.

One future line of inquiry is: what properties can (and what properties cannot) be efficiently recovered?  We now suggest how future work could address this question.

\section{Potential Networks}
\label{'section-potential networks'}
Our model can be conceptualized in a framework that we call ``potential networks".  Potential networks is a two phase model of social networks.  The first phase is the ``potential" network.  This network may not be directly observed or even exist an any normal manner.   The second phase is the ``behavioral" network, which is observable. However, the behavior network is realized by running some random process over the potential network which samples vertices and edges from it to produce the behavioral network and in some cases adds additional edges.

The key insight here is that we already have data from contagious networks and the process by which a contagious network grows acts ``locally"--ties are added to the community two people at a time.  Ethnographic tools could be used to build a model of how a particular technology spreads based on interviewing individuals.   Thus, this process may be much easier to observe than the original underlying network.  Then, based on this model, the data might be reverse engineered to recover ``global" properties of the underlying network.

Of course, real processes are more complicated than the models in this paper.  However, we do not think that this formidable challenge is insurmountable. By better modeling how particular contagious networks grow, we may be able to use the vast amounts of data to reconstruct properties of the underlying social network.


\paragraph{Related Work} \label{'paragraph potential related work'}

%
%

Recent work by Gomez-Rodriguez, Leskovec, and Krause~\cite{Gomez-RodriguezLK10} creates a model to try to infer a network of influence by looking only at the time sequence of an infectious outbreak (e.g. a news item through the blogosphere).   They show via computer simulations that their heuristics for recovering a potential network, given the timing data from a series of outbreaks, can simultaneously give high precision and recall of the original edges.  This model requires many cascades to be spread over the same nodes, while in the potential network setting, only one cascade is observed.

Questions similar to this have been looked at before in field of sampling theory, for example see \cite{ Cochran-77} and  \cite{ Sarndal-09}.  However, the techniques for reconstructing graph properties from sampled data is much smaller; see Chapter 5 in \cite{ Kolaczyk-09} for a survey of such results.  In fact, graphs have traditionally been hard to sample and this is part of the reason that the newly acquired large-scale data are so welcome.

Work by Handcock and Giles \cite{ Handcock-10} proposes a method of estimating properties from adaptively sampled networks by using maximum likelihood estimates over exponential graph models.   Moreover they show that their method does well on real test data.  A key observation made here is that simply because the sampled graph data is not representative of the graph, does not mean that key attributes cannot be reconstructed in a more clever way.  However this work seems to rely on an assumption that fails in our setting:  that which data are observed and which data remain unseen only depends on the observed data.  In a cascade, the fact that the cascade does not reach a person, already indicates that the person is likely not tightly connected to many infected nodes.

Finally, there has been recent work on how to sample a network in a way that makes it easy to recover certain properties (e.g.~\cite{LeskovecF06,StutzbachRDSW09}), however, in our framework the way that the network is sampled is fixed.

There are more results in sampling literature, but none seem to apply to the case where the part of the underlying graph that is observed depends on the underlying graph itself in a way that is not explicitly controlled by the sampler.

\section{Acknowledgments}
I would like to thank the many people whose help and insights have influenced this paper including Sarita Yardi, Matthew Salganik, Michael Mohoney, Jure Leskovec, and danah boyd.  Also, I would like state my appreciation for SNAP the Stanford Network Analysis Platform.

\bibliographystyle{abbrv}


\end{document}